\documentclass[9pt,twocolumn,twoside]{article}
\usepackage[utf8]{inputenc}
\usepackage{graphicx}
\usepackage{subcaption}
\usepackage{amsmath}
\usepackage{amssymb}
\usepackage{amsfonts}
\usepackage{tikz}
\usepackage{comment}
\usepackage{geometry}
\usepackage{float}
\usepackage{authblk}
\usepackage{hyperref}
\begin{document}
\title{RapidPIV: Full Flow-Field kHz PIV for Real-Time Display and Control}
\author[1]{Scott A. Bollt}
\author[1]{Samuel H. Foxman}
\author[1,*]{Morteza Gharib}

\affil[1]{GALCIT, California Institute of Technology, Pasadena, CA}

\affil[*]{sbollt@caltech.edu}
\maketitle

\begin{abstract}
We present a novel architecture for accelerating PIV calculations. An optical flow hardware accelerator does the brunt of the work, with cross-correlation only providing quick corrections. The result is RapidPIV: a free-to-download software program for real-time particle image velocimetry (PIV) for Linux and Windows computers with an Nvidia Turing-gen (or newer) GPU\footnote{RapidPIV download at \href{https://rapidpiv.caltech.edu}{https://rapidpiv.caltech.edu}}. Dense vector fields and vorticity can be displayed in real time when connected to compatible camera. Processing pre-existing PIV image files is also supported. We achieve 1,150 frames-per-second on 1 megapixel images. Accuracy, repeatability, and robustness are tested with physical experiments involving an accelerated flat plate and a cylinder wake. RapidPIV's accuracy, precision and ability to handle high displacement, velocity gradients, out-of-plane motion, and low seeding density compare favorably with the trusted multi-grid correlation software. However, RapidPIV is a thousand times faster.
\end{abstract}

\section{Introduction}
Particle Imaging Velocimetry (PIV) is a method by which the flow velocity field in a 2D or 3D domain may be measured by imaging tracer particles which passively advect with the flow. It is almost peerless in its ability to extract the fluid's entire hydrodynamic state (for homogeneous, incompressible flows). In its most basic form PIV works by illuminating a plane of particles suspended in a fluid with a laser sheet. Images of the particles' motion is recorded with a camera, and from these images, the motion of the fluid is inferred. 

With the introduction of digital methods, PIV became more repeatable and much less laborious because images could be transferred directly from digital cameras to digital computers where the images can be processed algorithmically. \cite{Willert1991} Digital PIV (DPIV) has since become ubiquitous in research as a flow diagnostic and a-posteriori measurement tool with a wide range of applications. 

Real-time PIV (RTPIV) is an implementation of PIV which is capable of producing vector fields from measurements with sufficiently little delay that the velocity fields are representative of the current state of the fluid. RTPIV expands the applicability of PIV in a few key ways. Because the velocity fields are being produced as the flow evolves, it becomes possible to integrate RTPIV into control schemes both in research and industrial settings. The flexibility and spatial resolution of PIV allows for sophisticated measurement and control schemes which are impractical or impossible with other kinds of sensors. The architecture and performance of RapidPIV makes it compatible with flow control, but the software itself does not support it. However, it does provide low-latency display of velocity fields as they are processed. That means it provides a visual display which is "real-time" to humans. 

RTPIV can also be used in situations where low latency is less important, because the throughput alone of RTPIV enables types of experiments or analysis which would simply be impossible otherwise. For instance, high throughput PIV processing with moderate latency (which RapidPIV is capable of) can enable expanded experimental campaigns either by reducing cycle time between experiments or increasing experiment runtime. Usually the processing time of PIV data exceeds the experiment time by orders of magnitude. For instance, recent experiments on the wakes of birds flying in formation used 148 recordings of 10 second 8MP video taken at 500Hz \cite{Hao2025}. From communication with the authors, processing this entire dataset takes 40 days of CPU time, so only a subset of the recorded data could actually be used for analysis, and it is a difficult and time consuming procedure to experiment with different setups and processing parameters. 

Various authors have contributed to the state of the art of RTPIV since its introduction in 1997. \cite{Arik1997} Indeed, a speedup exceeding an order of magnitude was achieved in the following two decades.\cite{Gautier2014,Wernet2019} Event-based camera methods have more recently come into favor and offer an alternative approach with impressive performance, but an entirely different set of trade-offs and performance measures.\cite{Drazen2011,Rusch2023} However, RTPIV has yet to see wide adoption even in research settings despite its theoretical and practical advantages. One of the major roadblocks to widespread adoption is the difficulty involved in setting up and implementing RTPIV. A RTPIV system requires delicately crafted code which, with exception of \cite{Munson2010}, is run on specialized hardware such as FPGAs, or more recently GPUs. Tight coupling to a video streaming camera is also required. For this reason, historically RTPIV systems have been built from the ground up by research groups around a specific camera and compute hardware. Furthermore, until recently few fluid systems of practical importance were within the range of proven RTPIV capabilities (as an exception to this rule, see \cite{Varon2019}). Regardless, for existing systems the fluid needs to evolve slow enough that dozens of frames per second can be considered real-time. Most existing RTPIV systems also cannot handle large displacements, large gradients, or most of the other challenges which naturally appear in typical PIV applications because they are based upon single-pass correlation with minimal pre and post-processing.

We address these challenges, in part, with RapidPIV: a software which can stream, from camera or disc, and process image data at over 1kHz. The result is a vast reduction in processing time and storage needs, as well as real-time PIV display/visualization. RapidPIV offloads the initial stage of the PIV workload to Nvidia's purpose-built optical flow hardware accelerator (NVOFA). \cite{Nvidiapatent} The NVOFA is a specialized circuit in all Nvidia GPUs starting with the Turing generation released in 2018. The sole purpose and capability of the Ada generation NVOFA is performing the memory-efficient Semi-Global Matching (eSGM) algorithm as quickly and power efficiently as possible. \cite{Hirschmuller2008} Because it was built to perform a single task, many electrical and logic optimizations are possible which are not possible on a general purpose computer. Camera communication, data handling, GUI, pre-processing, and post-processing code were developed in-house. The program is designed to be easy to use and as hardware-agnostic as possible (beyond needing an Nvidia GPU). The software works in Windows and Linux and is compatible with GenICam streaming cameras. The GUI allows users to view various real-time overlays at screen refresh rate such as velocity and vorticity (the program processes the data at full camera frame rate). Offline processing of image sequences with simultaneous display are also supported.

\section{RapidPIV}
RapidPIV is an app for Windows and Linux, with a CUDA-accelerated image processing pipeline centered around the NVOFA. RapidPIV is able to stream images from GenICam-compatible cameras, or load pre-existing particle image files. It calculates and displays processed velocity fields in real-time (up to the limit shown in figure \ref{syflow performance} and the screen refresh rate, respectively), and saves results to \texttt{.mat} files suitable for further analysis in Python or MATLAB. The GUI display is shown in the appendix, figure \ref{RapidPIV GUI}.

\begin{figure}[h!]
    \centering
    \includegraphics[width=0.75\linewidth]{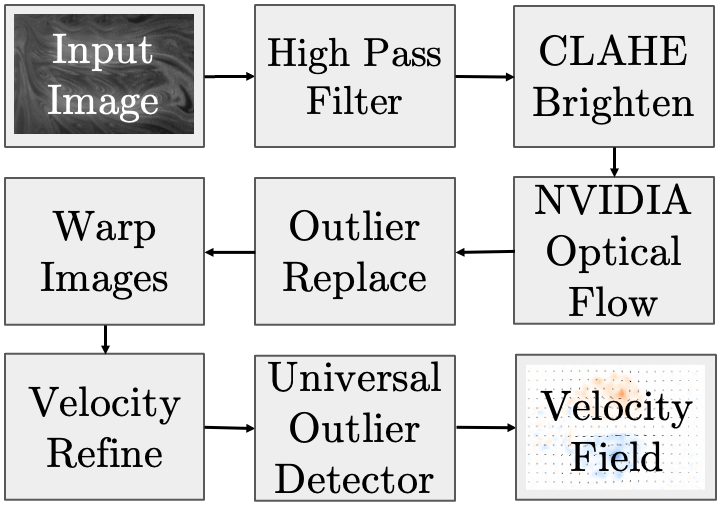}
    \caption{CUDA Processing Pipeline. RapidPIV filters and brightens images, calculates an initial velocity estimate with the Nvidia Optical Flow Accelerator, and refines the velocity field with classical PIV techniques.}
    \label{processing pipeline}
\end{figure}

In order to robustly generate precise and sub-pixel accurate velocity fields, RapidPIV pre-processes particle images before executing the NVOFA, and refines flow vectors afterwards with classical PIV techniques. The processing pipeline is shown in figure \ref{processing pipeline}. We now cover these stages in more detail.

First, images are grabbed as soon as available from the camera or folder, and uploaded to the GPU. To pre-process images, it applies a high-pass filter to remove static background, and performs contrast-limited adaptive histogram equalization (CLAHE) \cite{clahe} to brighten the image. Then, these images are sent to the NVOFA. The NVOFA calculates an estimate of the velocity field which is usually accurate to within ~0.5px. This is sufficient to act as an estimate of the velocity field, but is not accurate enough to be used for PIV analysis on its own. Any outliers are detected and removed. Then, exponential smoothing, with controllable smoothing constant $\alpha_1$, is applied to take advantage of the small temporal correlation of the NVOFA's random errors. 

Once the NVOFA velocity field has been refined as described above, RapidPIV warps the pre-processed particle image pair via central difference. From here, the processing is handled in much the same way as the standard PIV technique: Gaussian sub-pixel interpolation is performed on a 5-point correlation stencil about the shift which produces the maximum correlation coefficient (in this case the stencil's center is always chosen to be at zero-shift). This yields a sub-pixel correction. Then the universal outlier detector is applied (\cite{Westerweel2005}), and outliers are replaced with the median of their non-outlier neighbors. The last processing step is to perform exponential smoothing in time with smoothing constant $\alpha_2$ to take advantage of the low correlation between the sub-pixel velocity field estimates. The velocity fields (and optionally the processed images) can be saved via transfer from GPU to non-volatile storage (such as hard-drive), and/or displayed.

\section{Methods}
In this section we describe the experimental setups used to compare our software to a baseline, and the parameters used by RapidPIV and the baseline.

RapidPIV was tested using two different physical experiments against one of the best commercially available multi-grid PIV software programs, PIVview (we did not find the performance of free software sufficient to act as a baseline for these sequences). We chose a typical set of multi-grid parameters in PIVview as opposed to its most advanced features because the availability of advanced features varies between software programs. As a consequence, the results may not reflect the ultimate limit of what PIVview or its competitors are capable of, but we believe it indicates what a refined but standard multi-grid implementation is capable of. We therefore hereafter refer to the baseline as "multi-grid PIV." 
\subsection{Experiment: Plate Accelerated From Rest}
\begin{figure}[h!]
    \centering
    \includegraphics[width=0.75\linewidth]{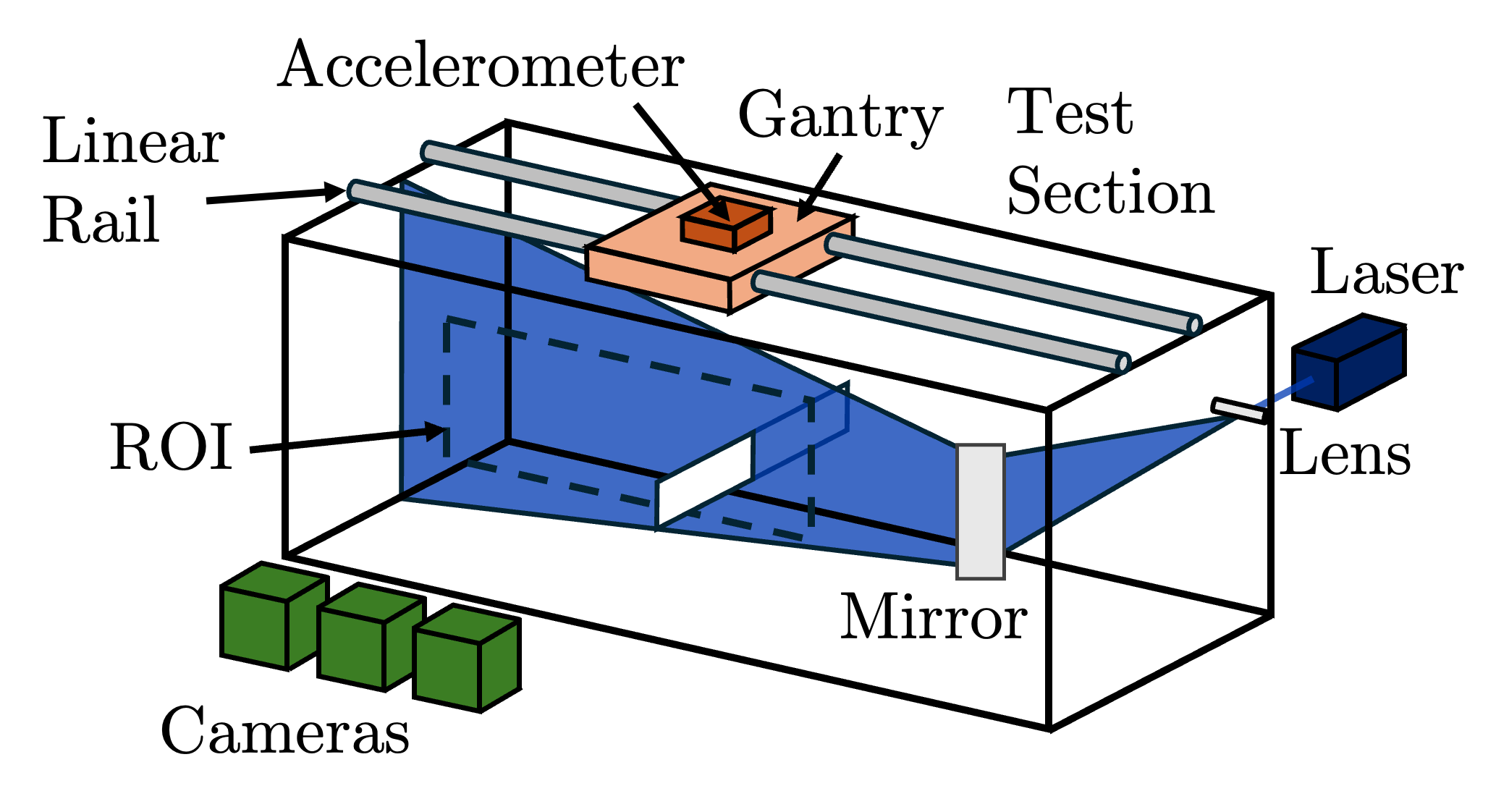}
    \caption{Perforated plate experimental setup. Endplates connecting plate to gantry not shown.}
    \label{perforated plate setup}
\end{figure}

A sharp-edged solid plate of chord length $c=61mm$ and width $315mm$ held between acrylic end-plates was accelerated along its normal at a constant rate from rest to a velocity of $v_f=0.075m/s$ through water ($Re\equiv\frac{cv_f}{\nu}=4900$) in a distance of $c/4$. The plate held this constant velocity up to a formation time of $T\equiv \frac{1}{c}\int_0^tv(\tau)d\tau=8.2$ before decelerating to rest. The flow was illuminated from behind at center-plane by a 450nm 10W laser expanded by a cylindrical lens. The plate is mostly transparent, except for the edges, so illumination is provided almost everywhere. The physical setup is shown in figure \ref{perforated plate setup}. Seeding was provided by 13$\mu m$ nominal diameter Potters Industries silver coated glass micro-spheres. A total of 1000 images, each, were taken at 100Hz via 3 synchronized IDT XSM-3520 High-Speed cameras aligned perpendicular to the laser plane. The images were unwarped and stitched together into a single 4999x1391px image such that no occluded region exists throughout the plate's motion. Intensity matching across the stitched region was enforced via contrast limited automatic histogram enhancement and high-pass spatial filter. Particle concentration is roughly 0.01-0.02ppp.

The processing options used for the multi-grid PIV method were: 16px final window size (uniformly weighted, 96x96px initial size), and 50\% stepover. Post-processing was provided by universal outlier detection (threshold 2, regularizer 0.1) and correlation-coefficient thresholding. Outliers were replaced via interpolation. All other settings were default. RapidPIV was evaluated using 16x16px correlation windows and 50\% stepover, with universal outlier detector threshold of 2, and universal outlier detector regularizer of 0.1px. The exponential smoothing was $\alpha_1=\alpha_2=0.5$.

This sequence was processed offline by RapidPIV. The flow contains a shear layer emanating from the plate's edges who's initial thickness is below what the seeding density can resolve. It breaks up into small yet distinct and resolvable vorticies via the Kelvin-Helmholtz instability. Most of the fluid is at rest, but the fluid in the Kaden spiral formed at startup is rotating and developing quickly early on, so velocity dynamic range and flow unsteadiness are high. Towards the end of the plate's motion the wake becomes turbulent and 3-dimensional. This makes loss of particle pairs substantial.
\subsection{Experiment: Cylinder Wake}
\begin{figure}[h!]
    \centering
    \includegraphics[width=0.75\linewidth]{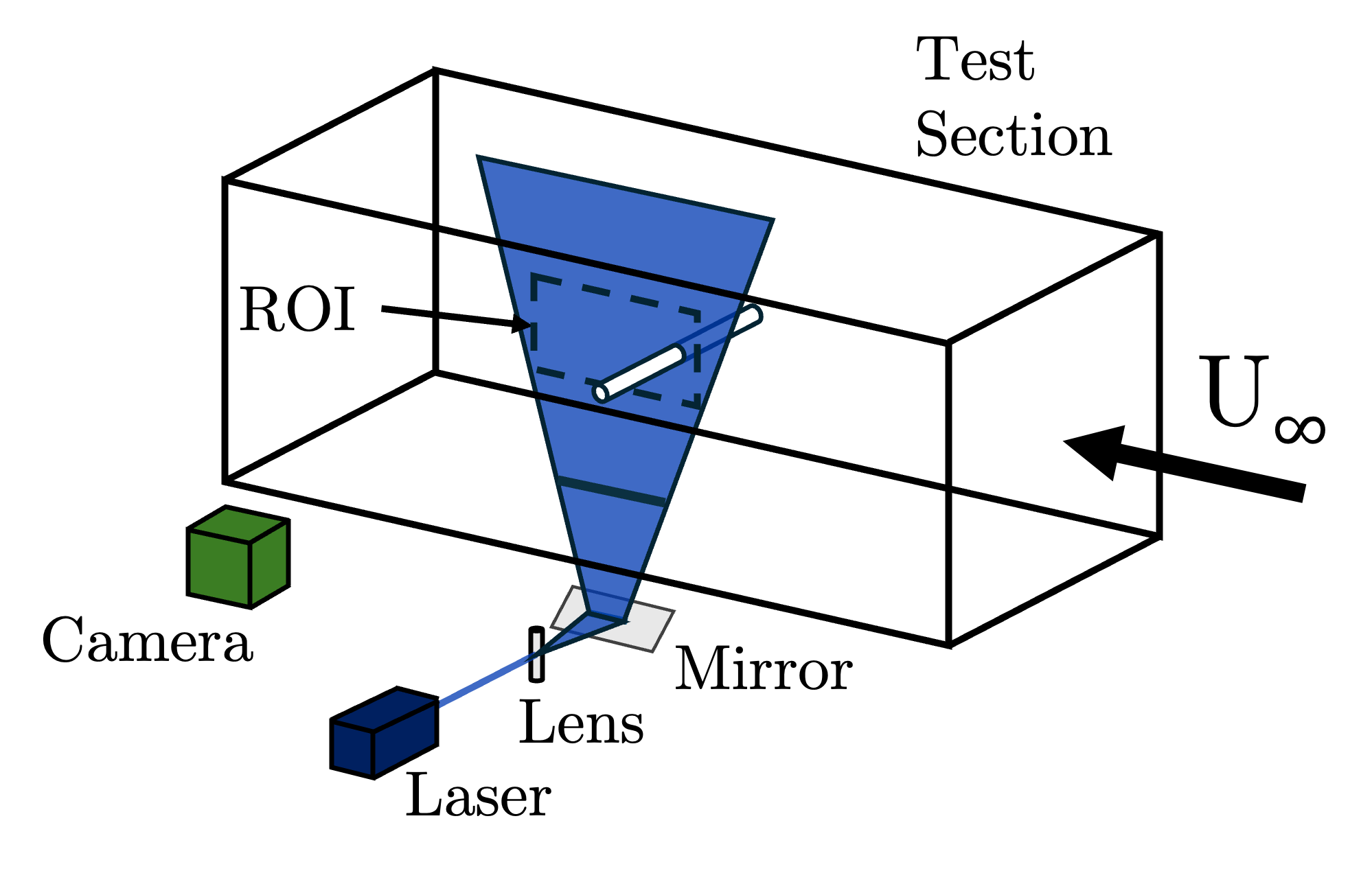}
    \caption{Cylinder experimental setup. Endplates not shown.}
    \label{cylinder setup}
\end{figure}

A cylinder of diameter $d=5.9mm$ and width $w=130mm$ was held between two end plates. The flow was illuminated at center-plane from below by the same laser. A flow of water ($Re\equiv\frac{u_\infty d}{\nu}=1300$) with velocity $u_\infty=0.020m/s$ was passed over it, as shown in figure \ref{cylinder setup} (for more information on the water channel used for this experiment, see \cite{Devey2025}). An Allied Vision Alvium U-158m camera was focused on the wake of the cylinder and took samples at 250fps. A total of 2500 image samples were acquired, with each image being 1456x544px. The particle concentration was roughly 0.002ppp, and the concentration of bright particles was substantially lower.

Because of the low particle concentration, resolving this flow is a challenge. Reference velocity fields were again provided by multi-grid PIV with the same options except the final window size was 32x32px and the initial window size was 160x160px. RapidPIV was evaluated at 32x32px resolution with 50\% stepover, with universal outlier detector threshold of 2, and universal outlier detector regularizer of 0.1px. The exponential smoothing was $\alpha_1=\alpha_2=0.5$.
\section{Results \& Discussion}
\subsection{Compute Performance}
\begin{figure}[h!]
    \centering
    \includegraphics[width=1\linewidth]{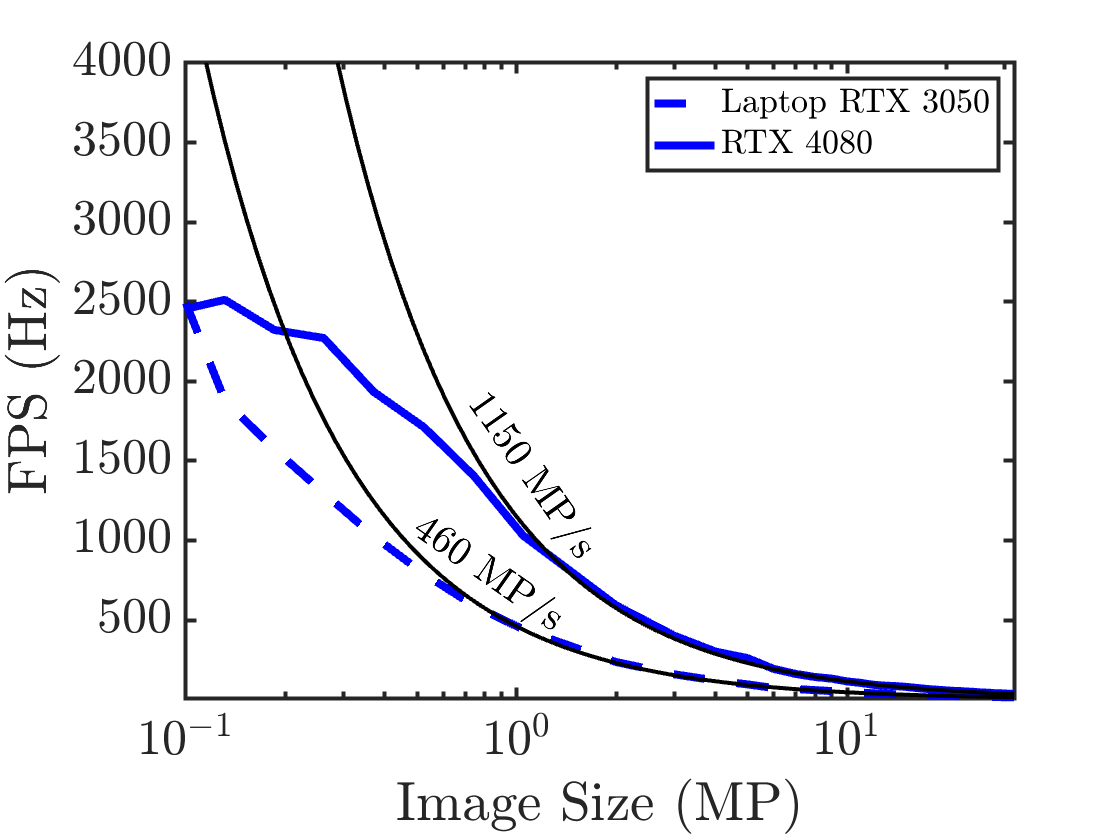}
    \caption{Log-linear plot of RapidPIV processing speed with Ada-generation (RTX 4080; solid blue line) and Ampere-generation (RTX 3050 Mobile Ti; dashed blue line) GPUs. Performance is measured in frames per second (vertical axis). The size of the images used for each data-point (in megapixels) is shown on a log scale on the horizontal axis.}
    \label{syflow performance}
\end{figure}

The compute performance of RapidPIV was tested on image sequences of various sizes up to the maximum it can handle: 8192x4096px, or 33.6 megapixels (MP). A laptop with Nvidia RTX 3050 Mobile Ti GPU (Ampere generation), and a desktop with Nvidia RTX 4080 (Ada generation) were used for the tests. The results, showing the number of full-density velocity field calculated per second at 50\% step-over and 16x16px window size, are shown in figure \ref{syflow performance}. The frame rate shown is the number of frame pairs processed per second, so if frame-straddling is used the camera's video capture rate would have to be twice the stated number to keep up with the computer. A few key trends emerge. First, both hardware options tested exceed prior performance capabilities. Second, for sufficiently small images the number of frames calculated per second (FPS) is independent of image size. We believe this is because for sufficiently small images the program becomes overhead limited. For larger images, roughly 1MP or more, the FPS becomes throughput limited; the frame-rate is inversely proportional to the size of the image. Indeed, for larger images RapidPIV's calculation speed can be calculated from the image size as,
\begin{equation}
    \text{FPS} = \text{(MP)}\cdot \text{(Throughput)},
\end{equation}
where $\text{Throughput}=1150\text{MP/s}$ for the RTX 4080, and $\text{Throughput}=460\text{MP/s}$ on the RTX 3050 Mobile Ti ($\approx 18$ million vecs/sec and 7 million vecs/sec, respectively). Interestingly, even on the largest image size supported, the RTX 4080 exceeds 30 FPS video rate. For comparison, multi-grid PIV measures at roughly 1MP/s on the desktop, over 1000 times slower on the same machine (the multi-grid PIV software uses the desktop's Intel i5-13600KF).

\begin{figure}[h!]
    \centering
    \includegraphics[width=1\linewidth]{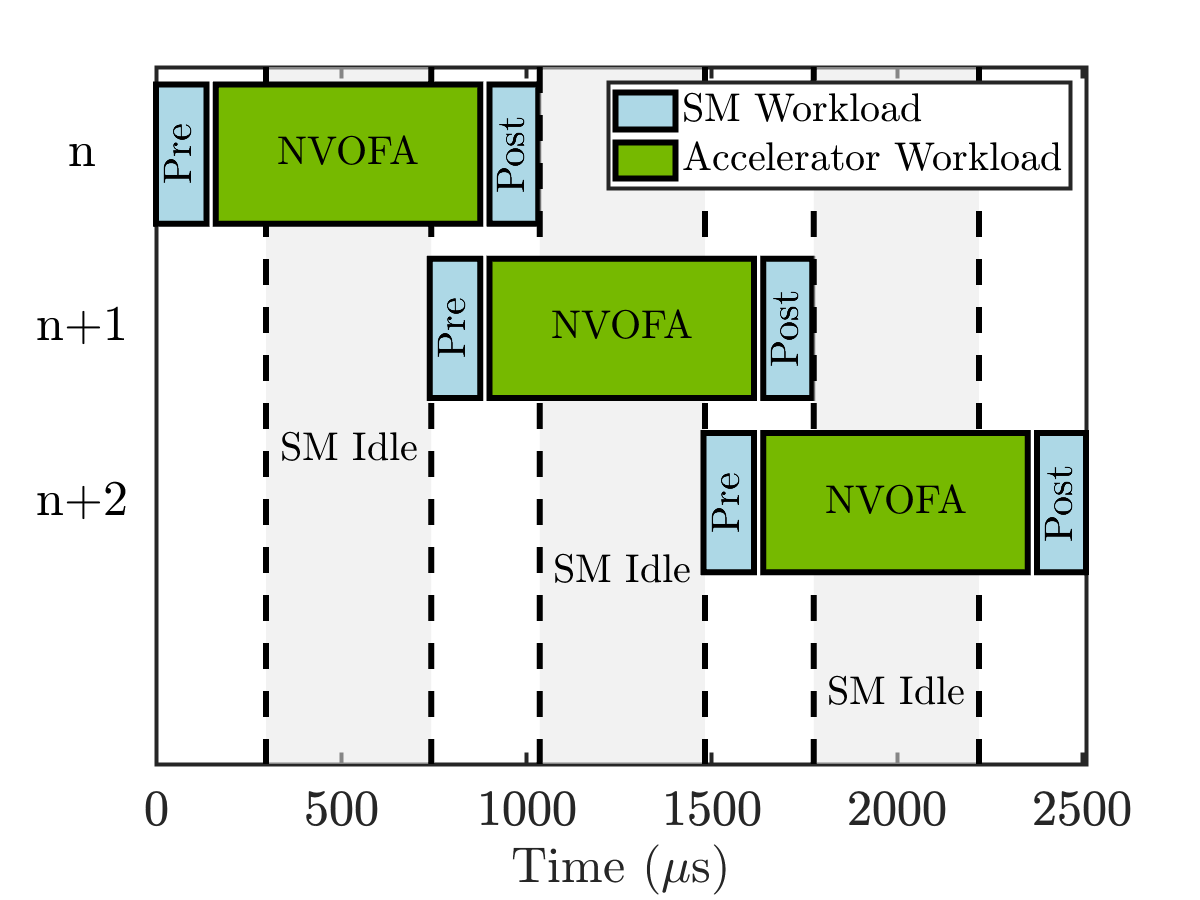}
    \caption{GPU Processing time breakdown for 1 MP images on RTX 4080. This figure is to scale. The pre and post-processing SM workloads have been each lumped to enhance readability. Latency of each frame is slightly more than what this figure would indicate, as this timing configuration is only possible if the camera and RapidPIV throughputs match exactly. As can be seen, the NVOFA is the limiting factor for the throughput. While the NVOFA calculates flow vectors, RapidPIV executes pre- and post-processing on the streaming multiprocessors (SMs), an area of the GPU independent from the NVOFA. The  kernels run very quickly, leaving idle time which could be used for further computation, such as more advanced PIV processing methods or separate machine learning tasks.}
    \label{syflow pipeline}
\end{figure}

The NVOFA is a separate resource, within the GPU, from the Streaming-Multiprocessors (SM's) which perform pre and post-processing. Figure \ref{syflow pipeline} shows, to scale, what GPU resources are used at each stage of the RapidPIV pipeline for 3 successive frames. Because the NVOFA is a separate resource to the SM's the post-processing on the $n$ frame and the pre-processing on the $n+2$ frame can take place while the NVOFA is running. Because the pre and post-processing together take less time than the NVOFA takes place to run, the pre and post-processing time do not effect throughput, only latency. This also means that the SM's are idle most of the time. There is therefore an intriguing possibility of using RapidPIV concurrently on the same GPU with other intensive processes, such as machine learning, or running more advanced pre or post-processing algorithms without performance tradeoffs.

On mobile devices, such as laptops or Nvidia Jetson modules, GPU power constraints can cause throttling and reduce throughput. For example, on a Dell XPS 9510 laptop, the RTX 3050 Ti Mobile GPU is limited to 40W, which throttles RapidPIV to 81\% NVOFA utilization when running the full processing pipeline. On the desktop machine with the RTX 4080, power consumption is 90W, far below its throttling limit of 320W. As a result, RapidPIV acheives 97\% NVOFA utilization. These differences contribute in part to the substantial performance difference seen in figure \ref{syflow performance} between these two pieces of hardware.

There are also possibilities associated with RapidPIV's compression capabilities, because when its source is a camera's video stream it becomes optional to save the raw image files. Processed velocity field \texttt{.mat} files of 1 MP images take 17.7 kB storage on average, at 50\% stepover and 16px window size, a 57x compression ratio (compared to the 1MB raw image files). In cases where it is acceptable to not store the raw images, this compression ratio enables continuous high-speed streaming of vector fields to hard-drive, side-stepping the RAM limitation on experiment run time of standard high-speed cameras (up to the throughput limit).
\subsection{Experiment: Accelerated Plate}
\begin{figure}[h!]
    \centering
    \includegraphics[width=\linewidth]{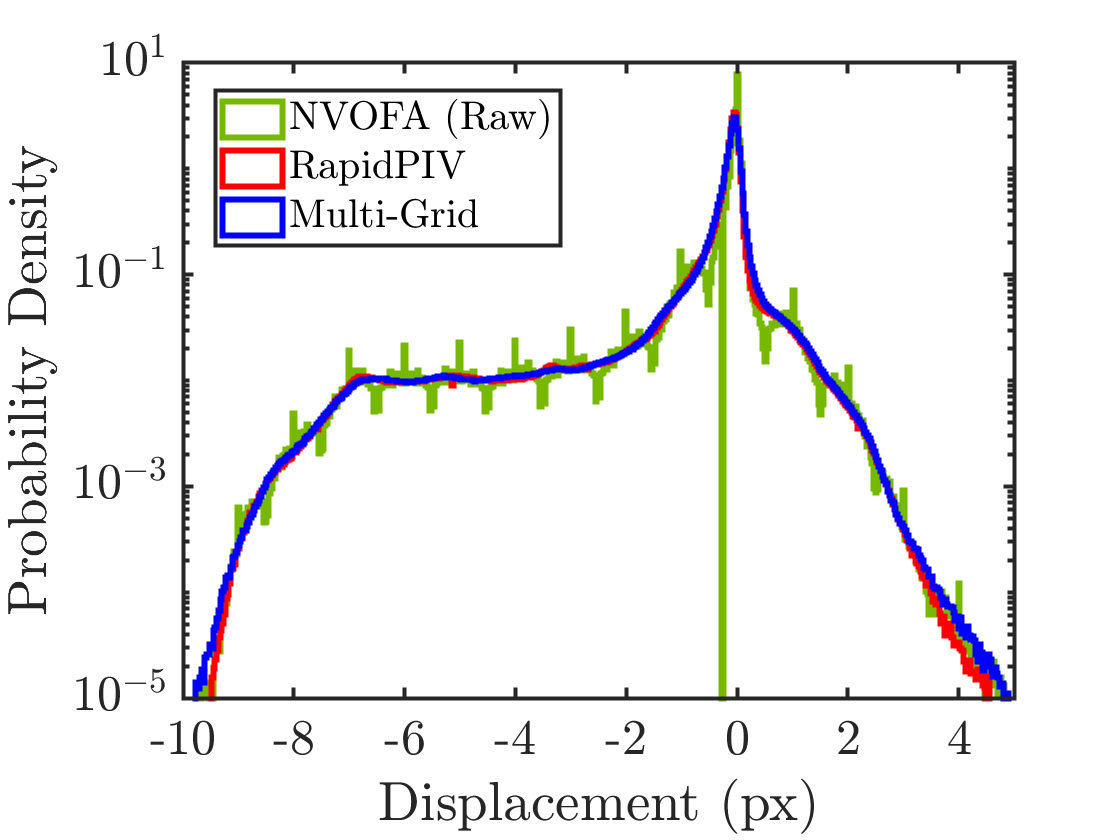}
    \caption{Probability density of the plate-normal flow velocity throughout the flow and throughout the experiment. The raw NVOFA (green) only matches the reference multigrid PIV distribution (blue) approximately. While it is sub-pixel accurate almost everywhere, it exhibits severe peak-locking. RapidPIV's pdf (red) matches the multigrid PIV pdf better.}
    \label{prob density plate normal}
\end{figure}

\begin{figure}[h!]
    \centering
    \includegraphics[width=\linewidth]{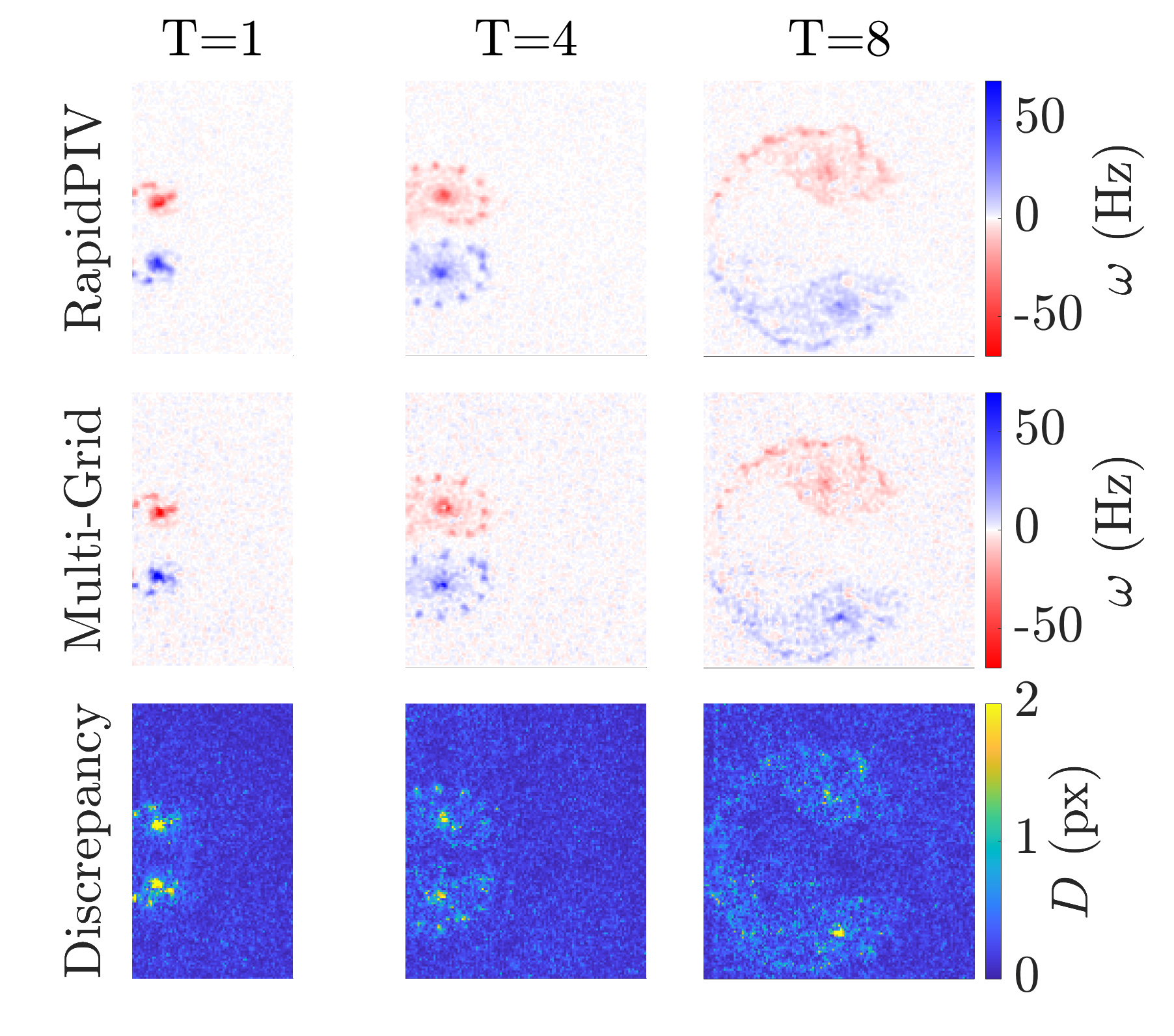}
    \caption{\textit{Upper and middle sub-plots:} Comparison of the vorticity field computed by RapidPIV (upper sub-plots) and multi-grid PIV (middle sub-plots) in the wake of a flat plate at various non-dimensional formation times, $T=\{1,4,8\}$. The vorticity fields correspond quite well, except the RapidPIV field appears to be a spatially smoothed version of the multi-grid PIV field: slightly less noisy and has fewer apparent artifacts but also slightly more smoothed in regions of high vorticity. \textit{Bottom sub-plots:} Velocity discrepancy between multi-grid PIV and RapidPIV at the same set of formation times as the sub-plots above. The discrepancy most everywhere is at the noise floor for multi-grid PIV, except for in regions of high vorticity: high vorticity regions, discrepancy can exceed 2 pixels.}
    \label{vorticity comparison plate}
\end{figure}

\begin{figure}[h!]
    \centering
    \includegraphics[width=\linewidth]{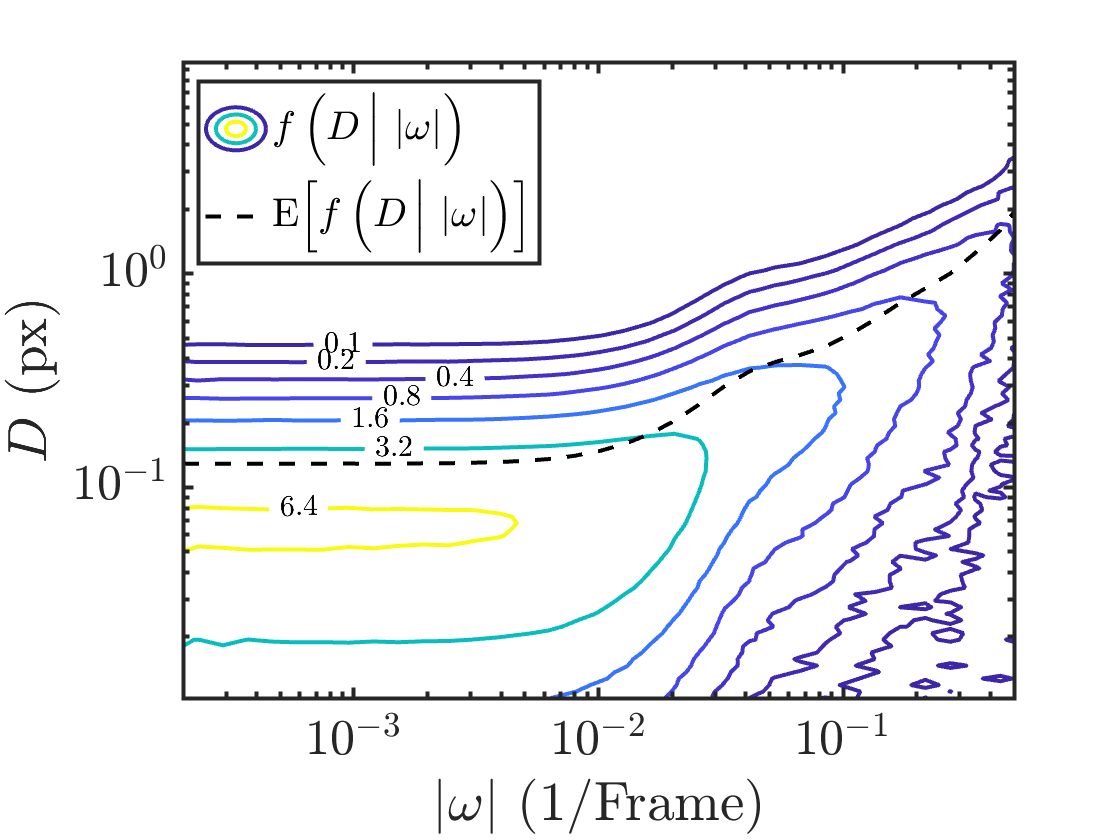}
    \caption{Log-log contour plot, with logarithmically spaced level sets, of the conditional probability distribution of the discrepancy between RapidPIV and multi-grid PIV (conditioned on the magnitude of vorticity), as a function of the magnitude of vorticity: $f\left(D\, \Big | \,|\omega|\right)$. This distribution was obtained by sampling the whole flow, except the edges, throughout the plate's movement. The expectation value is also shown (black $--$). When $|\omega|$ is small, $D$ is small and independent of the vorticity distribution indicating the discrepancy here is explained by their combined noise-floors. For larger $|\omega|$, the discrepancy grows with $|\omega|$ indicating one or both of the software develop bias in highly vortical regions, as expected.}
    \label{discrepancy correlation}
\end{figure}

\begin{figure}[h!]
    \centering
    \includegraphics[width=\linewidth]{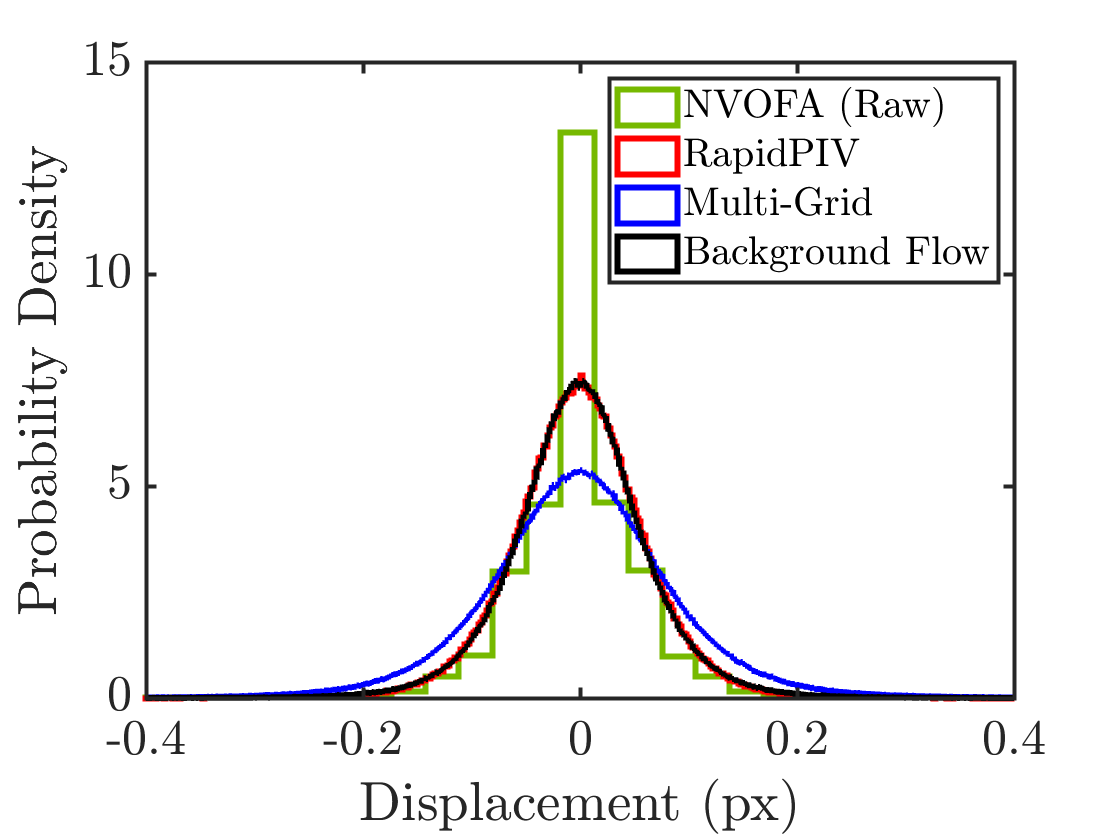}
    \caption{Probability density function of the image displacement between frames, before the plate begins motion. The background turbulence intensity in the nominally "quiescent" fluid results in a roughly 0.1px-per-frame standard deviation in particle displacement (black curve). The distribution of the displacement rate as measured by the raw NVOFA (green curve), RapidPIV (red curve), and multi-grid PIV (blue curve) all mostly capture this background turbulence intensity level. However, RapidPIV comes the closest followed by PIVlab and finally the raw NVOFA (the NVOFA's wider bins reflect its 1/32px displacement resolution).}
    \label{zero_flow_noise}
\end{figure}

The plate impulsive start experimental data showed a generally excellent agreement between RapidPIV and multi-grid PIV. Consider first the probability density functions (pdf's) of the plate-normal fluid displacement between frames as measured by the raw NVOFA, RapidPIV, and multigrid PIV: $f(\underline{v}_{{NVOFA}}\cdot \hat{e_1})$, $f(\underline{v}_{{Rapid}}\cdot \hat{e_1})$, $f(\underline{v}_{{multi}}\cdot \hat{e_1})$. These probability densities are shown together in figure \ref{prob density plate normal}. It is clear from $f(\underline{v}_{{NVOFA}}\cdot \hat{e_1})$ that the raw NVOFA is insufficient for most PIV applications: peak locking and discretization error are present. Meanwhile, $f(\underline{v}_{{Rapid}}\cdot \hat{e_1})$ is the same as $f(\underline{v}_{{multi}}\cdot \hat{e_1})$ aside from slightly lower noise (narrower pdf around $0$) at the cost of slight under-prediction of the velocity tails in log-scale.
Consider the discrepancy field between RapidPIV and multigrid PIV,
\begin{equation}
    D \equiv \underline{v}_{{Rapid}} - \underline{v}_{{multi}}.
\end{equation}
Figure \ref{vorticity comparison plate} shows that $D<0.2px$ most places in the flow, which according to figure \ref{zero_flow_noise} is consistent with the combined random errors of multi-grid PIV and RapidPIV. However, RapidPIV appears to apply slightly more spatial smoothing to the vorticity field than multigrid PIV does. Partially for this reason, $D$ is higher in highly vortical regions. However, this is not an indication per-se that RapidPIV is under-performing. Indeed, close inspection of the vorticity field at $T=4$ indicates what appears to be more artifacts in the multi-grid PIV field than the RapidPIV field. In particular, the upper starting vortex as measured via multi-grid PIV has a region of near-zero vorticity just off-center, while the lower one has an unusually sharp vorticity profile at the center. These artifacts are fairly confined, but they do not appear in RapidPIV, nor do they appear when multi-grid PIV is used with the advanced feature, correlation-plane summing (not shown).
The relationship between $D$ and the magnitude of vorticity, $|\omega|$, can be represented by the conditional probability of $D$ given $|\omega|$: $f\left(D\, \Big | \,|\omega|\right)$. This conditional, as a function of $|\omega|$, is shown in figure \ref{discrepancy correlation}. For small $|\omega|$, discrepancy is small, however for $|\omega|>10^{-2}$ larger vorticities tend to correlate with larger discrepancies. For the largest vorticities (which are rare, hence the higher noise in this region of the PDF) discrepancies can exceed 2 pixels (The max discrepancy across the entire field and the entire sequence is 5px).
The vorticity-invariant low vorticity magnitude behavior of $f\left(D\, \Big | \,|\omega|\right)$ suggests a noise floor. To compare the random error of the raw NVOFA, multi-grid PIV, and RapidPIV, the pdf of displacement components was calculated for each of these methods in a well illuminated region of the fluid behind the plate, for the first 100 frames (before the plate was set in motion). The result of this calculation is displayed in figure \ref{zero_flow_noise}. The background flow was determined using the pdf of the temporal mean of the multi-grid PIV velocity field (the flow varied little during those 100 frames, so the temporal mean is justified). The fluid itself was not perfectly quiescent, so an ideal histogram should match the one of the background flow, not be a delta function. The NVOFA histogram peak-locked onto $0$ so it under-predicted the turbulence intensity. RapidPIV slightly over-predicted turbulence intensity due to random error, while the larger random error of multi-grid PIV resulted in a larger over-prediction. This supports our previous observation that RapidPIV produces lower noise because it slightly spatially smooths the velocity fields (which is undesirable), and we are doing a small amount of temporal filtering via the exponential average.
\subsection{Experiment: Cylinder Wake}
\begin{figure}[h!]
    \centering
    \includegraphics[width=\linewidth]{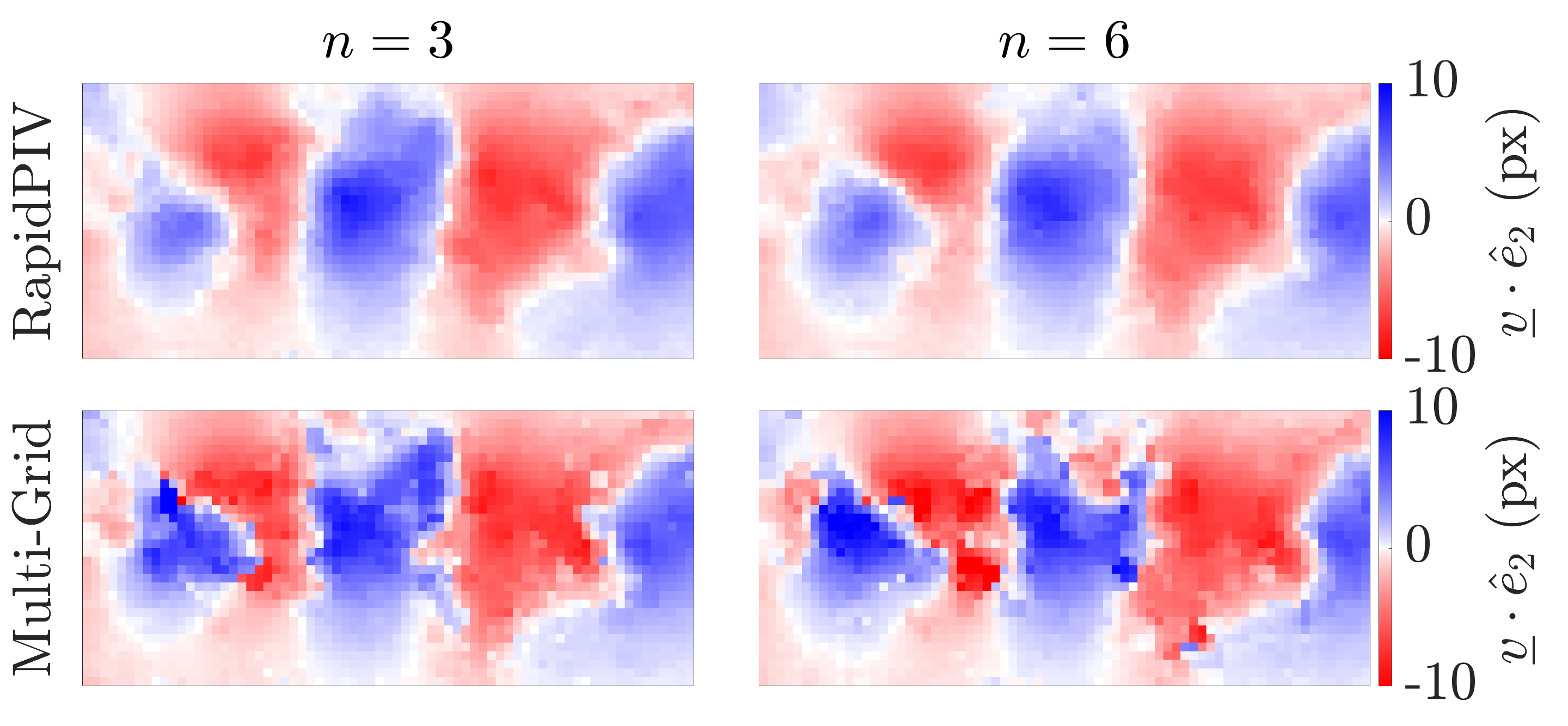}
    \caption{Comparison of the vertical velocity in the Von-Karman shedding wake of a circular cylinder at two near-by times (frame 3, and frame 6) as calculated using RapidPIV and multigrid PIV. The frames are spaced far enough in time that the temporal filtering in RapidPIV cannot explain the higher level of consistency in the RapidPIV frames compared to the multigrid PIV frames, but close enough that the flow has not evolved appreciably. The multigrid PIV sequence has outlier regions that change randomly frame-to-frame throughout the entire 2499 vector-field sequence. These regions do not appear to exist in this sequence.}
    \label{cylinder vertical velocity}
\end{figure}

\begin{figure}[h!]
    \centering
\includegraphics[width=0.4\linewidth]{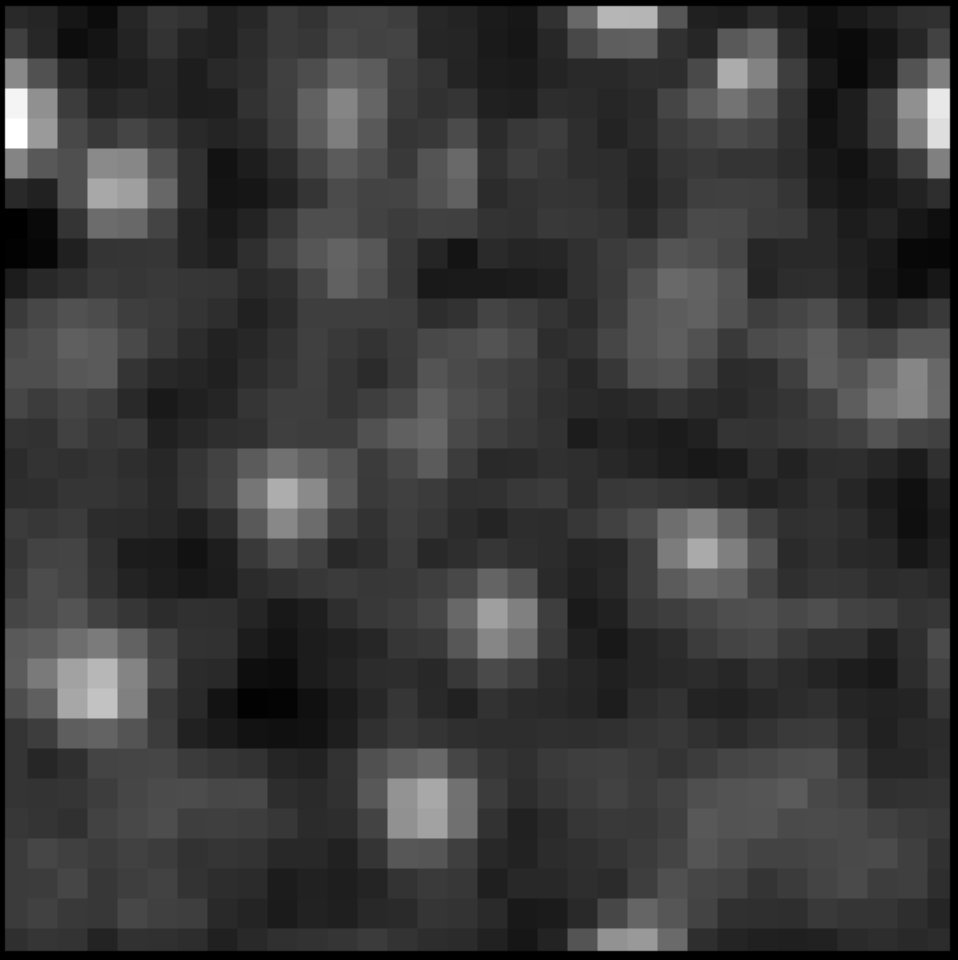}
    \caption{Correlation plane from multi-grid PIV at a point where the vector is an outlier. There are many candidates for the peak associated with the true flow displacement, which an exhaustive search could falsely identify. Because RapidPIV relies on the NVOFA rather than exhaustive search to determine the location where peak interpolation is performed, the existence of other peaks is irrelevant to RapidPIV's robustness (assuming the NVOFA did its job).}
    \label{cylinder correlation plane}
\end{figure}

The cylinder wake experiment presented a challenge to multi-grid PIV whereas RapidPIV could reliably obtain velocity fields at a window size of 32x32px (except at the edges of the flow where neither method performed because of out-of-frame motion, not shown). The sequence is really a better candidate for PTV than PIV since the seeding density is so low compared to the size of the velocity gradients, however it provides an interesting point of comparison for how the two methods tackle the problem of velocity estimation. Figure \ref{cylinder vertical velocity} shows the vertical velocity in the wake of the cylinder, as measured by RapidPIV and multi-grid PIV, at two frames tightly spaced in time. The flow has evolved little in the elapsed 0.012 seconds (the flow has advected 0.04 cylinder lengths), so the scalar field should be nearly the same. This is observed to be largely true of RapidPIV, but it is not true of multi-grid PIV: groups of artifacts are present in the multi-grid PIV velocity field in both frames. To avoid these artifacts in multi-grid PIV a lower final resolution is required.
To understand why RapidPIV outperforms multi-grid PIV in this scenario, we must look to the correlation plane. Figure \ref{cylinder correlation plane} shows the correlation plane extracted from multi-grid PIV at a point where the velocity vector is an outlier. There are many equally good candidates for the correct flow displacement. If by chance the correct peak is not the highest one, as is the case here, an outlier results from exhaustive search of the correlation plane. Even though RapidPIV also relies on correlation, the peak which gets sub-pixel approximated is determined apriori by the NVOFA, not by exhaustive search. If the NVOFA provides an approximation which is close enough, the 5-point correlation stencil will refine the true peak rather than a spurious one. In that case, the fact that there are other equally high correlation peaks elsewhere in the correlation plane is of no consequence. There are so few particles per flow structure that the larger scale velocity fields used to refine the search at smaller scales, in the multi-grid method, are not useful at the smaller scales. This dataset is therefore likely a better candidate for particle tracking, but the fact that the NVOFA can use neighboring regions to interpolate gracefully where there is missing data helps it perform outside the realm of where PIV can usually be applied.

\section{Conclusions and Future Work}
We developed an architecture which enables high-speed PIV processing on GPU's. This architecture is packaged into a flexible, PIV processing program with real-time velocity display capability. It is capable of either processing pre-existing image sequences, or continuously streaming from GeniCam cameras, into non-volitile memory (ie, the hard-drive). This was achieved by utilizing Nvidia's optical flow hardware accelerator to quickly produce a velocity field estimate. This estimate, while somewhat crude, is sufficiently close to the true velocity field that classical PIV techniques could be used to refine the field. Accuracy, precision, and robustness of the system were tested on experimental data. Precision was found to be comparable to standard multi-grid PIV techniques. For accuracy, the velocity produced by RapidPIV agreed with multi-grid PIV almost everywhere. The exception is highly vortical regions where both software begin to fail in different ways. Robustness was often similar, but the cylinder experiment suggests that RapidPIV can perform better than multi-grid PIV in cases where particle tracking is better suited than both. 

The processing speed of RapidPIV is roughly 1000 times faster than the multi-grid code we used (on the same desktop, although the multi-grid code does not take advantage of the GPU). When only the velocity fields are saved, just 1/57th of the storage space is required, and since RapidPIV is compatible with streaming cameras, it becomes possible to only save the vectors. These properties could allow this tool to expand the uses of PIV into areas where PIV was previously too slow, expensive, or impractical. For instance: biological or turbulence experiments which require exceedingly long run times at high frame rates, exploratory PIV-based flow visualization, or PIV for freely configurable scalar sensing. In the avian experiment mentioned in the introduction (see \cite{Hao2025}), RapidPIV would allow each dataset to be processed in just 40 seconds rather than 7 hours, allowing data processing to keep up with the experiments so that bird handling rather than PIV processing is the most time consuming step. This would allow rapid experimental iterations which are currently impossible. The technique of combining optical flow hardware-accelerators with classical PIV algorithms may also yield advancements in RTPIV for control applications, as it both greatly improves throughput, and leaves plenty of GPU resources free for other uses. These resources could be utilized to simultaneously execute computationally intensive control schemes, such as those based on machine learning. Low-power optical flow accelerators exist, which could enable onboard PIV applications. 

In closing, we would like to point out that experiments at high Reynolds numbers have been and continue to be much more time efficient than solving for the flow numerically, but the experimentalists ability to utilize this advantage depends on the data they can extract in a timely manner.\cite{Gharib1996} PIV is one of our best tools for matching or exceeding the resolution of simulations, however the processing of PIV data has always been time consuming. Whether it was particle tracking by hand on film in the days of yore, or clever processing of images taken on digital cameras (both of ever-greater performance), extracting the velocity field has always been by far the limiting step. Thus processing has, in a way, prevented us from unlocking the full capability of flow facilities. We hope this program will help alleviate the processing bottleneck. Also, in developing this software, we have discovered an entirely different method by which velocity fields may be calculated at up to roughly 1,000,000 MP/s, in principle without loss of accuracy or robustness. This would make imaging rather than processing the limiting step because the fastest streaming cameras are currently limited to roughly 25,000 MP/s, and the existing PCIe 5.0 standard could only support two of them due to its 63GB/s bandwidth (event cameras could elevate the PCIe bottleneck, but not until the technology catches traditional cameras). There is therefore the question of what can be done with such capability. We call upon the scientific community to answer the following question: what can be done with PIV if the real-time speed limit is set by the camera, no matter what camera?
\section{Acknowledgments}
We would like to thank Luiz Lourenco and IDT for providing the XSM-3520 High-Speed cameras for the plate experiment, and Kenneth Breuer and Siyang Hao for discussions on the PIV requirements for avian experiments.
\section{Funding}
Support was provided in part by the Center for Autonomous Systems and Technologies (CAST) at Caltech.

This material is based upon work supported in part by the National Science Foundation Graduate Research Fellowship under Grant No. DGE‐1745301. Any opinions, findings, and conclusions or recommendations expressed in this material are those of the author(s) and do not necessarily reflect the views of the National Science Foundation.

\section{Disclosures}
A patent has been filed in relation to this work.

\bibliographystyle{apalike}

\newpage
\section{Appendix}
\begin{figure}[h!]
    \centering
    \includegraphics[width=\linewidth]{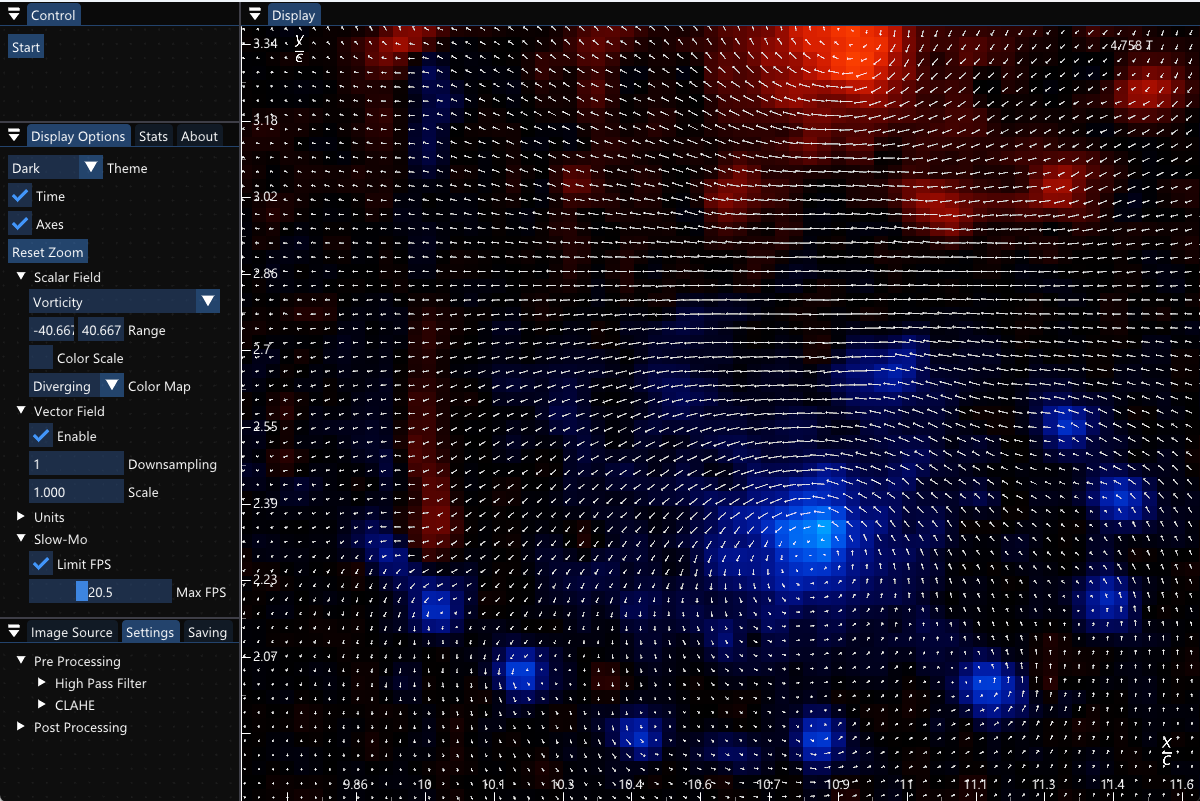}
    \caption{Example of the RapidPIV GUI. Velocity and vorticity fields are overlaid using the available options. The control options are on the left side. The flow has been zoomed into a field of interest via scroll-to-zoom and panning with the mouse. The native processing speed of this sequence is roughly 200fps (roughly 7MP images) which is too fast for the human eye to follow, especially when zoomed in. Therefore, the "slow-mo" option has been activated to set playback speed at 20.5fps. Dimensionless position is shown via axes on the edges, and the dimensionless time is displayed in the top right corner.}
    \label{RapidPIV GUI}
\end{figure}
\end{document}